# Some Approaches to the Energy-Loss Straggling Calculation

P.B. Kats*[1], A.V. Kudravets*, A.S. Rymasheuskaya*, O.O. Voskresenskaya**[2]

*Brest State A.S. Pushkin University, 21 Kosmonavtov Boulevard, Brest, 224016, Belarus*

*\*\* Joint Institute for Nuclear Research, 6 Joliot-Curie St., Dubna, Moscow Region, 141980 Russia*

Some exact and approximate methods commonly used to calculate corrections to the Bethe stopping formula are modified and adapted by the authors to the calculation of the energy loss straggling (ELS). An intercomparison is carried out for results of approaches developed in the present work. An excellent agreement obtained between the results for ELS, calculated with an exact method previously proposed by one of the authors and the results of the Lindhard–Sørensen method for moderate relativistic energies. It is shown that some modified by authors approximate methods for the ELS calculating have the higher accuracy than the conventional approximate method of Lijian−Qing−Zhengming (LQZ). So, in all the cases considered, the accuracy of the twice modified LQZ method in the ELS calculations is higher than the accuracy of the usual LQZ method. A thrice modified LQZ method was also suggested in this work. A number of shortened LQZ methods for the ELS calculating have been proposed and developed too.

## 1. Introduction

Energy loss straggling (ELS) denotes the development of the width and shape of the energy loss spectrum of an initially monochromatic beam as a function of time or pathlength. Straggling is an inherent feature of stopping measurements which cannot be reduced indefinitely by making more measurements. It has atomistic and statistical aspects [Sigmund, 2004, 2006]. The study of the ELS of charged particles is of significant interest not only because of their fundamental aspects but also because of their impact on many applications in physics and other sciences. So, accurate information on ELS is important in the application of the Rutherford backscattering spectrometry, nuclear reaction analysis, and scanning transmission ion microscopy. As ELS is one of the main factors limiting depth resolution, understanding the shape of energy loss distribution rather than energy distribution is important in the application of particle identification. In view of these applications, the study of ELS in matter has become an interesting topic in recent years [Lohmann *et al.*, 2023; Margin *et al.*, 2022; Dib *et al.*, 2021; Mery *et al.*, 2021; Quisim *et al.*, 2021; Selau *et al.,* 2021, 2020; Duy *et al*., 2020; Novikov and Teplova, 2020; Nguyen *et al*., 2019; Khodyrev, 2019; Mahalesh, and Basavaraja, 2018; Kumar

---

[1]*E-mail addresses:* katspyotr@yandex.ru
[2]*E-mail addresses:* voskr@jinr.ru

*et al.*, 2018; Devendrappa and Sannakki, 2018; Damache *et al.*, 2017; Babu and Sjue *et al.*, 2017; Babu *et al.*, 2016; Badiger, 2016, Sigmund *et al.*, 2014, 2008].

The outline of this papers is as follows. Section 2 considers adaptation the rigorous (Lindhard−Sørensen, Voskresenskaya et al.) methods for calculating the ELS, comparison of the results obtained by these both methods for $Z > 0$ (Sections 2.1−2.4) and $Z < 0$ (Section 2.5), as well as their relative corrections to the first Born approximation. Section 3 presents approximate Lijian−Qing−Zhengming (LQZ) method (Section 3.1) and its twice and thrice modifications (Section 3.2) for calculating the ELS of heavy nuclei due to close collisions. Intercomparison of accuracy of all these methods for calculating the ELS for elements from uranium to ohaneson were performed in Section 3.3. In Section 4 coefficients for a series of shortened variants of the LQZ method (LQZS methods) are calculated, formulas for the ELS calculating by various shortenes LQZS methods are obtained, the velocity-averaged relative errors are estimated for different LQZS methods and their accuracy analyzed for $Z = -118$ (Section 4.1) and $Z = -60$ (Section 4.2). Section 5 summarizes our results and conclusions. Appendix A shows the derivation of the basic formula for $X$ required to calculate the ELS. Appendix B contains the proposed trice-modified LQZ-method.

## 2. Rigorous methods for calculating the ELS of heavy nuclei due to large momentum transfers

The contribution to the fluctuation of heavy charged particles comes from small and large transfers of momentum to the electrons of matter [Scheidenberger *et al.*, 1996]. Further in this work, only the case of large momentum transfers will be considered.

### 2.1 Method of Lindhad−Sørensen in application to the ELS calculation

According to [Lindhard and Sørensen, 1996; Sigmund, 2006], the average square fluctuation in energy loss (variance of energy loss/dispersion/energy loss straggling) during the passage of particles through matter can be defined as

$$W = \langle (\delta E - \langle \delta E \rangle)^2 \rangle, \qquad (1)$$

where $\delta E$ is the energy loss. This fluctuation is entirely dependent on the close collision. For relativistic particle nuclei, taking into account close collisions per unit path in matter, the increase in the average square fluctuation can be described by the formula

$$\frac{dW}{dx} = 4\pi Z^2 e^4 n Z' \gamma^2 X \qquad (2)$$

with gamma factor $\gamma = 1/\sqrt{1-\beta^2}$, relative velocity $\beta = v/c$ of decelerating particle with charge number $Z$, the atomic number $Z'$ of the atom of the decelerating substance, the concentration of atoms in the substance $n$, and the dimensionless quantity $X$ equal to 1 for Rutheford scattering.

Accounting for close collisions with matter electrons for a point nucleus in the framework of nonrelativistic quantum mechanics leads to the so-called Bohr formula [Feldman *et al.*, 1986].

$$\frac{dW}{dx} = 4\pi Z^2 e^4 n Z'.$$

As for the $X$ value, then in the first-order quantum perturbation theory the Dirac equation leads to

$$X = 1 - \beta^2/2. \qquad (3)$$

However, it has been experimentally proven [Scheidenberger *et al.*, 1996] that the use of the first Born approximation for heavy relativistic ions leads to underestimated results.

We will to obtain the $X$ value for the relativistic case to calculate the ELS within the Lindhard−Sørensen transport theory by a direct calculation analogous to that the transport fluctuation cross section, which directly connected to the increase in energy loss fluctuation (see Appendix

The result will be applicable, among other things, to finite nuclear sizes, which must be taken into account for ultrarelativistic nuclei. In this case, the $X$ value can be represented by the following series:

$$X = \frac{2}{\eta^2} \sum_{k=-\infty}^{\infty} |k| \Big[ \frac{k-1}{2k-1} \sin^2(\delta_k - \delta_{k-1}) + \frac{1}{2(4k^2-1)} \sin^2(\delta_k - \delta_{-k}) -$$
$$- \frac{(k-1)(k-2)}{2(2k-1)(2k-3)} \sin^2(\delta_k - \delta_{k-2}) - \frac{k-1}{2(2k-3)(4k^2-1)} \sin^2(\delta_k - \delta_{-k+1}) - \quad (4)$$
$$- \frac{k+1}{4(2k+1)} \left( \frac{1}{4k^2-1} + \frac{1}{4(k+1)^2-1} \right) \sin^2(\delta_k - \delta_{-k-1}) \Big].$$

$$\eta = Z\alpha/\beta.$$

For a point nucleus, the Coulomb phase shifts have the form:

$$\delta_k = \frac{1}{2i} \ln\left[\frac{k - i\eta/\gamma}{\rho_k - i\eta}\right] - \arg\Gamma(\rho_k + 1 + i\eta) - \frac{\pi}{2}(\rho_k - l_k),$$

$$\rho_k = \sqrt{k^2 - (Z\alpha)^2}, l_k = \begin{cases} k, k > 0 \\ -k-1, k < 0, \end{cases} \quad (5)$$

where $\Gamma$ is the gamma function.

### 2.2 Scheidenberger et al. approach to the ELS determination

For moderate relativistic energies, where scattered bare nuclei can be regarded as point charges, their close-collision result for the energy loss straggling, according to [Scheidenberger *et al.*, 1996], can be retrieved with the exact Mott cross section:

$$\frac{d\text{W}}{dx} = 2\pi n Z' \int_{\theta_0}^{\pi} E^2(\theta) \frac{d\sigma_M}{d\Omega} \sin\theta d\theta. \quad (6)$$

Here, $E(\theta)$ is the energy transferred to the target electron in a collision leading to scattering through an angle $\theta$ in the center of mass system; $\theta_o$ is the angle corresponding to the minimum transferred energy at which it can be considered that the nucleus collides with free electrons (it is usually assumed in calculations that $\theta_o = 0$, because for $\theta_o \ll 1$, the value of the integral depends weakly), and $d\sigma_M/d\Omega$ denotes the exact Mott cross section with the scattering cross section solid angle $\Omega$.

The Mott exact scattering cross section $d\sigma_M/d\Omega$ from (6) expressed in terms of conditionally convergent series in Legendre polynomials, so direct calculation of the integral (6) is difficult. In this regard, the "relativistic Bohr formula", based on the first Born approximation for the Mott scattering cross section, is often used to the ELS calculation. But, as noted [Scheidenberger *et al.*, 1996; Voskresenskaya, 2017], the use of the first Born approximation for heavy ions leads to an underestimated ELS value compared to the experimental data.

### 2.3 Voskresenskaya et al. approach to the ELS calculation

In [Voskresenskaya, 2017] was obtained an another than (4) X representation based on the Mott exact cross section in the form of quite rapidly converging series of quantities bilinear in the Mott partial amplitudes [Voskresenskaya et al., 1996], which can be easily computed. In [Voskresenskaya, 2017, 2018] this approach was adopted to calculating the so-called central moments:

$$\mu_{n,M} = 2\pi n Z' \Delta x \int_0^\pi [E(\theta)]^n \frac{d\sigma_M}{d\Omega} \sin\theta d\theta,$$

where $\Delta x$ is the thickness of the layer of matter traversed by the ion. The second central moment in this case is the dispersion of energy loss.

If we use the $X$ value introduced above, then, according to [Voskresenskaya, 2017], with the notation of [Kats *et al.*, 2021] we get an expression for $X$ in the form of fast convergent series:

$$X_V = \frac{2}{\eta^2} \sum_{k=0}^{\infty} (k+1) \left[ \left(\frac{\eta}{\gamma}\right)^2 \left| \frac{F_M^{(k)}}{2k+1} - \frac{F_M^{(k+1)}}{2k+3} \right|^2 + \left| \frac{kF_M^{(k)}}{2k+1} - \frac{(k+2)F_M^{(k+1)}}{2k+3} \right|^2 \right], \qquad (7)$$

where

$$F_M^{(k)} = \frac{i}{2}(-1)^k [kC_M^{(k)} + (k+1)C_M^{(k+1)}], \quad C_M^{(k)} = -e^{-i\pi\rho_k} \frac{\Gamma(\rho_k - i\eta)}{\Gamma(\rho_k + 1 + i\eta)}. \qquad (8)$$

The therms of this series decrease like

$$l^{-2n+1} (n = 2, 4), \text{ when } l \to \infty.$$

As shown in [Voskresenskaya, 2017], the application of the corresponding expressions based of this new representation of the exact Mott cross section leads to results that are consistent with the experiment [Scheidenberger *et al.*, 1996].

### *2.4 Comparison of the rigorous methods to the ELS calculation*

The formula for the ELS calculating obtained in [Voskresenskaya, 2017] is less cumbersome than that obtained in [Linhard and Sørensen, 1996]. Therefore, it is of interest to compare the results of ELS calculating using formulas (4) and (7), which has not yet been done.

For comparison, we calculated by analogy with (12) in [Voskresenskaya, 2017] the following quantity, which characterize the relative correction of the Born approximation in computing the value $X$:

$$\delta = \frac{X - X_B}{X_B}. \qquad (9)$$

For calculations in the sum in (4), the upper and lower limits of 3000 and –3000, respectively, were set, and 3000 was taken as the upper limit in (7). The results are given in Table 1.

**Table 1**
Relative correction to the first Born approximation $\delta$ on the data of works [Lindhard and Sørensen, 1996] [1] and [Voskresenskaya, 2017] [2] for $\beta = 0.75$ (1) and $Z = 92$ (2).

| $\beta = 0.75$ | | | | | |
|---|---|---|---|---|---|
| Z | 10 | 20 | 30 | 40 | 50 |
| [1] | 0.0486257 | 0.117467 | 0.210810 | 0.333171 | 0.489170 |
| [2] | 0.0486257 | 0.117467 | 0.210810 | 0.333171 | 0.489170 |
| Z | 60 | 70 | 80 | 90 | 100 |
| [1] | 0.683263 | 0.919165 | 1.19858 | 1.51831 | 1.86377 |
| [2] | 0.683263 | 0.919165 | 1.19858 | 1.51831 | 1.86377 |
| $Z = 92$ | | | | | |
| $\beta$ | 0.05 | 0.15 | 0.25 | 0.35 | 0.45 |
| [1] | 0.0174471 | 0.138572 | 0.333661 | 0.561661 | 0.801533 |
| [2] | 0.0174471 | 0.138572 | 0.333661 | 0.561661 | 0.801533 |
| $\beta$ | 0.55 | 0.65 | 0.75 | 0.85 | 0.95 |
| [1] | 1.04791 | 1.30526 | 1.58614 | 1.91352 | 2.32962 |
| [2] | 1.04791 | 1.30526 | 1.58614 | 1.91352 | 2.32962 |

Thus, when using the limits $N = 3000$, the results of calculations according to [Lindhard and Sørensen, 1996] and [Vosrresenskaya, 2017] coincide up to the seventh decimal digit over the range of approximately $1 \leq \gamma \leq 15$.[3]

The relative difference between $X$ values for $Z = 6$ and $\beta = 0.999$

$$\delta X = \frac{X - X_V}{X_V} \cdot 100\%$$

shows Figure 1:

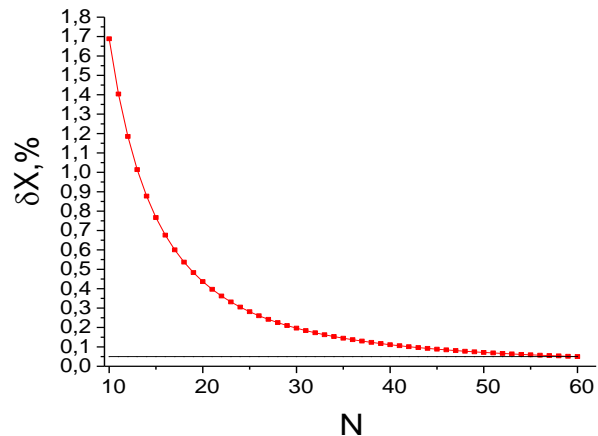

**Fig. 1.** Relative difference between the $X$ values calculated from the data of [Lindhard and Sørensen, 1996] and [Voskresenskaya, 2017] at $Z = 6$ and $\beta = 0.999$.

---

[3] This result can be understood, since the correction of Lindhard and Sørensen can be regarded as the sum of the Mott and Bloch corrections, but obtained in a different way (see [Kats *et al.*, 2021]).

Here, *N* are the limits in the sums of expressions (4) and (7). Already at *N* = 60, the difference becomes less than 0.05%. For a given velocity and large *Z*, or for a given *Z* and lower velocity, this difference is reached even earlier.

The calculation time for ten values of the relative correction according to [Lindhard and Sørensen, 1996] was 37 seconds, and according to [Voskresenskaya, 2017] it was 15 seconds. Therefore, and also because of the simpler expression, in our opinion, it is preferable to use the method developed in [Voskresenskaya, 2017].

Formula (2) can be also represented in terms of the normalized Mott cross section (NMCS) [Kats *et al.*, 2022]:

$$\frac{dW}{dx} = 2\pi n Z' \int_{\theta_0}^{\pi} E^2(\theta) \frac{d\sigma_R}{d\Omega}(1-\beta^2) R_{KHV}(\theta) \sin\theta d\theta,$$

$$\frac{d\sigma_R}{d\Omega} = \left(\frac{Ze^2}{2mv^2}\right)^2 \frac{1}{\sin^4(\theta/2)}. \qquad (10)$$

Considering that the energy transferred to an electron in a collision is related to the scattering angle in the center-of-mass system by the expression

$$E(\theta) = \frac{2mc^2\beta^2}{1-\beta^2}\sin^2\left(\frac{\theta}{2}\right), \qquad (11)$$

it is easy to obtain an expression for *X* in terms of the NMCS :

$$X = 0.5\int_0^{\pi} R_{KHV}(\theta)\sin\theta d\theta. \qquad (12)$$

### 2.5 *Some features of the exact methods for calculating the ELS of heavy nuclei*

Let us consider the case *Z* < 0. This may correspond to the motion of antinuclei in matter or nuclei in antimatter. Table 3 presents the *X* values calculated by both of the above exact methods. It demonstrates that for both negative and positive *Z*, both methods give the same results for the fluctuation of energy loss. Further, we will use the method [Voskresenskaya, 2017], as the expression for *X* in this method is simpler and computer summation of series is achieved faster.

**Table 2**

Comparison of the X value obtained by methods of [Voskresenskaya, 2017] [1] and [Lindhard and Sørensen, 1996] [2] for $\beta = 0.75$ and $Z = -92$.

$\beta = 0.75$

| Z   | −10      | −20      | −30      | −40      | −50      |
|-----|----------|----------|----------|----------|----------|
| [1] | 0.695479 | 0.681311 | 0.673983 | 0.671573 | 0.672504 |
| [2] | 0.695479 | 0.681311 | 0.673983 | 0.671573 | 0.672504 |
| Z   | −60      | −70      | −80      | −90      | −100     |
| [1] | 0.675537 | 0.679736 | 0.684431 | 0.689168 | 0.693664 |
| [2] | 0.675537 | 0.679736 | 0.684431 | 0.689168 | 0.693664 |

$Z = -92$

| $\beta$ | 0.05     | 0.15     | 0.25     | 0.35     | 0.45     |
|---------|----------|----------|----------|----------|----------|
| [1]     | 0.998811 | 0.989442 | 0.970471 | 0.941426 | 0.901211 |
| [2]     | 0.998811 | 0.989442 | 0.970471 | 0.941426 | 0.901211 |
| Z       | 0.55     | 0.65     | 0.75     | 0.85     | 0.95     |
| [1]     | 0.847772 | 0.778335 | 0.690093 | 0.580741 | 0.448687 |
| [2]     | 0.847772 | 0.778335 | 0.690093 | 0.580741 | 0.448687 |

In [Voskresenskaya, 2017, 2018] there are graphs of the dependence of quantity (9) at a fixed velocity ($\delta = \delta(Z)$) and at a fixed Z ($\delta = \delta(\beta)$). It turned out that for $Z = 92$ and $\beta = 0.95$ the quantity $\delta$ reaches a value of 2.33, i.e., the exact result exceeds the result in the first Born approximation by a factor of 3.33. We considered the Z interval from −118 to 118 at $\beta = 0.75$ and the velocity interval from 0.05 to 0.99 c for $Z = 118$. The results illustrate Figures 2a, 2b.

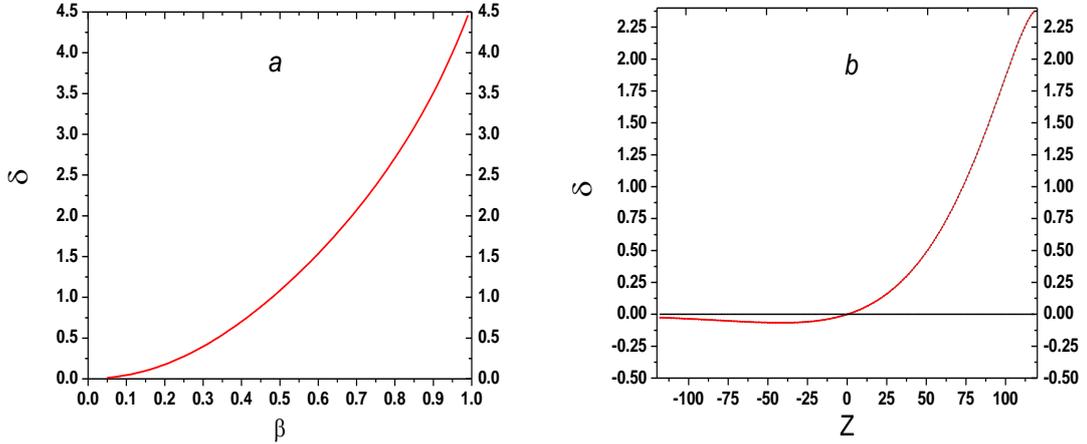

**Fig. 2.** Relative correction (9) to the first Born approximation for X at $Z = 118$ (*a*) and $\beta = 0.75$ (*b*).

It follows from Figure 2a that at $Z = 118$ and sufficiently high velocities $\delta(\beta)$ approaches the value 4.5. As $\beta \rightarrow 1$, the quantity $\delta(\beta)$ approaches the value 4.6 that is the exact energy loss fluctuacion is almost 5.6 times higher than that obtained in the first Born approximation.

However, at very high velocities, it is already necessary to take into account the size of the nucleus, which is impossible in the approach based on the exact Mott cross section.

From Figure 2b, the following conclusions can be drawn:

- the relative correction $\delta(Z)$ is negative for negative $Z$;
- the modulus $\delta(Z)$ is much less for large negative $Z$ than for its positive values;
- the value of the function $\delta(Z)$ does not change monotonically with a change in $Z$.

Similar results were obtained for the Mott correction in [Kudravets, 2022]. The Mott correction is introduced when calculating the average energy losses in matter to take into account the difference between the exact Mott cross section and that obtained in the first Born approximation.

The dependence $\delta(\beta)$ for $Z = -118$ is shown in Figure 3. Since $\delta(\beta)$ is small for negative $Z$, the result is given as a percentage. This implies that the relative correction $\delta(\beta)$ is not only nonmonotonic for negative $Z$, but can also change sign. The maximum positive $\delta(\beta)$ value for $Z = -118$ is $\delta(\beta) = 0.40\%$ at $\beta = 0.456$.

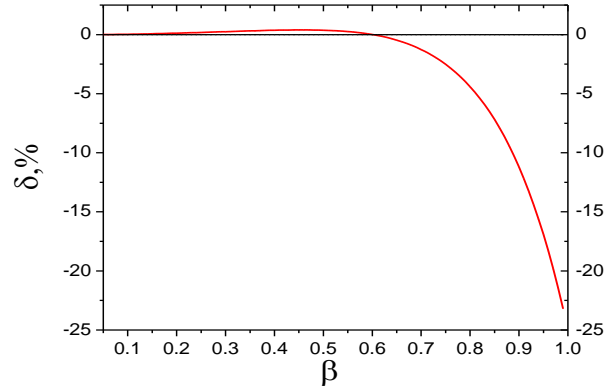

**Fig. 3.** Relative correction $\delta(\beta)$ (9) to the first Born approximation for $Z = -118$.

The formula for calculating $X$ can be reduced to the form (12) with the normalized Mott scattering cross section $R$:

$$R = \frac{d\sigma_M}{(1-\beta^2)d\sigma_R}, \quad \frac{d\sigma_R}{d\Omega} = \left(\frac{Ze^2}{2mv^2}\right)^2 \frac{1}{\sin^4(\theta/2)}, \quad (13)$$

where $d\sigma_R / d\Omega$ is the Rutherford scattering cross section.

Figure 4 shows the exact normalized Mott cross section (NMCS) and its first Born approximation [Mott, 1932] as functions of the angle $\theta$ for $\beta = 0.456$ and $\beta = 0.999$ at $Z = -118$.

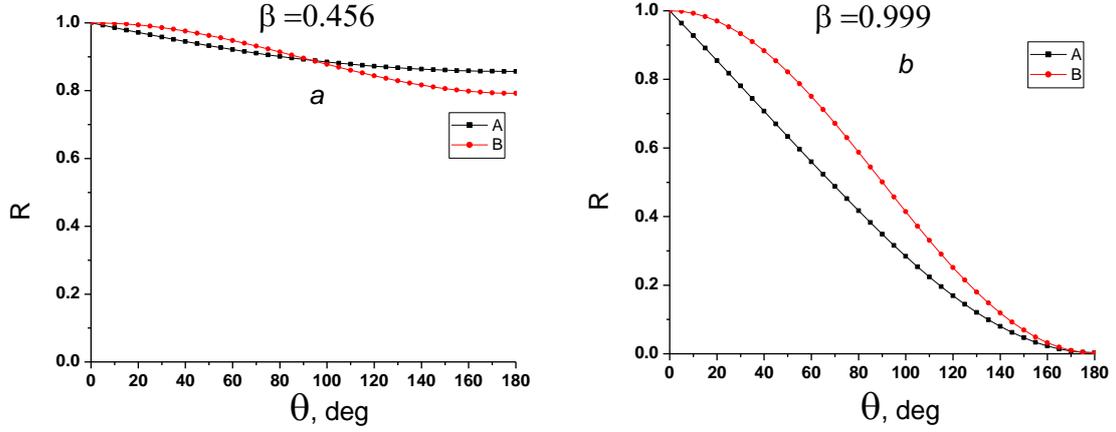

**Fig. 4.** The exact normalized Mott cross section (A) and its first Born approximation (B) as fuctions of the angle $\theta$ for $Z = -118$ at $\beta = 0.456$ (*a*) and $\beta = 0.999$ (*b*).

For $\beta = 0.456$ at large $\theta$, the exact Mott cross section becomes larger than the cross section in the first Born approximation, which results in $X > X_B$ and $\delta > 0$. For scattering at $\beta = 0.999$, the exact Mott cross section is smaller than the cross section in the first Born approximation that leads to $X > X_B$ and $\delta < 0$.

The Mott correction, obviously, for any $\beta$ and $Z < 0$ does not turn out to be greater than 0, because the contribution to it comes from small angles at which the exact Mott cross section for $Z < 0$ is smaller than the cross section in the first Born approximation.

Let us consider the dependence $\delta(Z)$ for $\beta = 0.456$ (Figure 5). For a given velocity, the sign of $\delta(\beta)$ is also not constant for negative $Z$.

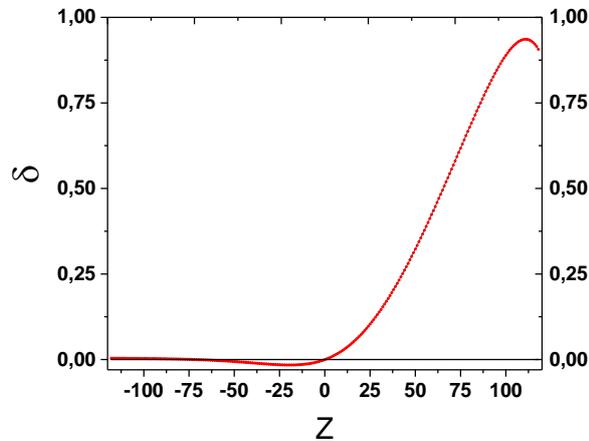

**Fig. 5**. Relative correction $\delta(Z)$ to the first Born approximation for $X$ at $\beta = 0.456$.

It is interesting that quantities δ, and hence X, turn out to be nonmonotonic functions of Z for positive Z. A similar nonmonotonicity is also observed for the Mott correction (Figure 6). In [Morgan and Eby, 1973], where Mott's correction was considered as a function of Z and β, such a feature was not revealed.

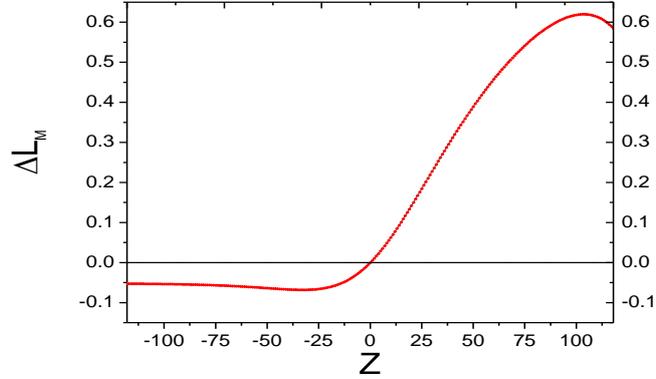

**Fig. 6.** The Mott correction as a function of Z for the β value 0.456.

In conclusion, we note that the value of $XZ^2$, which determines the dispersion of energy loss, as shown by the calculation, increases monotonically with increasing modulus Z.

## 3. Approximate methods for calculating the ELS of heavy nuclei due to close collisions

### 3.1 Approximate Lijian-Qing-Zhengming method

One of the most accurate methods for calculating the Mott cross section is the approximate LQZ method [Lijian *et al.*, 1995] that also can be adapted to the calculation of ELS.

These authors have suggested the following approximation of the NMCS:

$$R_{LQZ}(\theta; Z, \beta) = \sum_{j=0}^{4} a_j(Z, \beta)(1-\cos\theta)^{j/2},$$

$$a_j(Z, \beta) = \sum_{k=1}^{6} d_Z(j,k)(\beta - \bar{\beta})^{k-1}, \quad \bar{\beta} = 0.7181287. \tag{14}$$

The coefficients calculated for each Z value. They are given in [Lijian *et al.*, 1995] for Z = 1–90 and in [Boschini *et al.*, 2013] for Z = ± (1–118). This NMCS form makes it easy to find analytical expressions for various integrals. In contrast to the Mott correction, for which the integral diverges when the lower limit is set to zero, no singularit arises in this case.

Substitution (2) into (4) and elementary integration leads to the result:

$$X_{LQZ} = a_0 + \frac{2\sqrt{2}}{3}a_1 + a_2 + \frac{4\sqrt{2}}{5}a_3 + \frac{4}{3}a_4. \tag{15}$$

To characterize the accuracy of the LQZ method, we will use, as in the second section, the quantity

$$\delta X_{LQZ} = \frac{X_{LQZ} - X_V}{X_V}. \tag{16}$$

The values $\langle |\delta X_{LQZ}| \rangle$ averaged over 26 velocities (from 0.1 c to 0.999 c) are presented in Table 3. For comparison, we present also the NMCS average relative error over velocities and angles:

$$\langle ER \rangle_{LQZ} = \frac{1}{26}\sum_{i=0}^{26} ER_{LQZ}(\beta_i),$$

$$ER_{LQZ} = \sqrt{\frac{\sum_{i=0}^{36}[R_{LQZ}(\theta_i;Z,\beta) - R_{KHV}(\theta_i;Z,\beta)]^2}{\sum_{i=0}^{36} R_{KHV}(\theta_i;Z,\beta)^2}}, \tag{17}$$

where $R_{KHV}(\theta;Z,\beta)$ is the NMCS calculated by the method considered in [Kats et al., 2022].

**Table 3**
Relative errors of NMCS and modulus of X for Z > 0 (1) and Z < 0 (2), averaged over velocities ($\langle |\delta X_{LQZ}| \rangle$) and angles ($\langle ER \rangle_{LQZ}$).

1. $\langle |\delta X_{LQZ}| \rangle$,% and $\langle ER \rangle_{LQZ}$,% for $Z>0$

| Z | 6 | 12 | 26 | 36 | 52 | 60 |
|---|---|---|---|---|---|---|
| $\langle |\delta X_{LQZ}| \rangle$ | 1.71·10⁻³ | 1.42·10⁻² | 2.24·10⁻² | 1.92·10⁻² | 8.40·10⁻² | 0.103 |
| $\langle ER \rangle_{LQZ}$ | 4.11·10⁻³ | 1.91·10⁻² | 9.85·10⁻² | 0.177 | 0.281 | 0.347 |
| Z | 75 | 80 | 92 | 104 | 114 | 118 |
| $\langle |\delta X_{LQZ}| \rangle$ | 0.105 | 0.102 | 8.16·10⁻² | 0.145 | 0.268 | 0.332 |
| $\langle ER \rangle_{LQZ}$ | 0.693 | 0.883 | 1.43 | 1.94 | 3.09 | 4.09 |

2. $\langle |\delta X_{LQZ}| \rangle$ and $\langle ER \rangle_{LQZ}$ for $Z<0$

| Z | −6 | −12 | −26 | −36 | −52 | −60 |
|---|---|---|---|---|---|---|
| $\langle |\delta X_{LQZ}| \rangle$ | 1.96·10⁻³ | 3.39·10⁻³ | 1.15·10⁻² | 9.74·10⁻³ | 9.04·10⁻³ | 1.39·10⁻² |
| $\langle ER \rangle_{LQZ}$ | 2.90·10⁻³ | 7.60·10⁻³ | 0.023 | 0.023 | 0.024 | 0.023 |
| Z | −75 | −80 | −92 | −104 | −114 | −118 |
| $\langle |\delta X_{LQZ}| \rangle$ | 1.17·10⁻² | 8.77·10⁻³ | 3.93·10⁻³ | 9.68·10⁻³ | 1.24·10⁻² | 1.24·10⁻² |
| $\langle ER \rangle_{LQZ}$ | 0.013 | 0.010 | 0.012 | 0.020 | 0.022 | 0.022 |

As can be seen from the Table 3, the indicated errors in the *X* calculation are generally less than the errors of the NMCS. For large *Z*, they can differ by an order of magnitude or more. For *Z* < 0, these errors are of the order of a percent or less.

The LQZ method [Lijian *et al.*, 1995] can be applied for the simplified ELS calculations,. It is used, for example, in the FLUKA program for modeling elementary particle interactions [Battistoni *et al.*, 2016].

## 3.2 Modified versions of LQZ method for calculation of energy loss straggling of nuclea due to close collisions

In [Kats *et al.*, 2022] a twice modified LQZ method (LQZ$_{m2}$) for NMCS calculation has been developed. This method gives generally high accuracy for calculating the NMCS and Mott's correction (see [Kats *et al.*, 2022]). In this work, we develop a thrice modified method for calculation of the NMCS (see Appendix B). In this section, we adapt these modified methods to compute the ELS. For this purpose, we will obtain analytical expressions for calculating *X* within the LQZ$_{m2}$ and LQZ$_{m3}$ methods and check their accuracy.

As can be seen from the Table 3, for *Z* < 0 the calculation error for *X* is on the order of $10^{-2}$ % and less. The high accuracy of the LQZ method for calculating NMCS at *Z* < 0 was noted also in [Boschini *et al.*, 2013]. This makes it possible to develop simplified "shortened" versions of the LQZ method to calculate *X* at *Z* < 0 (see Section 4).

As for the modified LQZ$_{m2}$- and LQZ$_{m3}$-methods, in them coefficient $a_0$ is replaced by 1 in the expression for *R*, and terms of the form $a_5(1-\cos\theta)^{5/2}$ and $a_5(1-\cos\theta)^{5/2} + a_6(1-\cos\theta)^3$ are added to the LQZ$_{m2}$ and LQZ$_{m3}$ methods, respectively. Taking into account (12), we calculate the integrals:

$$0.5\int_0^\pi a_5(1-\cos\theta)^{5/2}\sin\theta d\theta = \frac{8\sqrt{2}}{7}a_5, \tag{18}$$

$$0.5\int_0^\pi a_6(1-\cos\theta)^{6/2}\sin\theta d\theta = 2a_6. \tag{19}$$

Adding the corresponding terms to (15), we obtain expressions for *X*:

$$X_{LQZm2} = a_0 + \frac{2\sqrt{2}}{3}a_1 + a_2 + \frac{4\sqrt{2}}{5}a_3 + \frac{4}{3}a_4 + \frac{8\sqrt{2}}{7}a_5, \qquad (20)$$

$$X_{LQZm3} = a_0 + \frac{2\sqrt{2}}{3}a_1 + a_2 + \frac{4\sqrt{2}}{5}a_3 + \frac{4}{3}a_4 + \frac{8\sqrt{2}}{7}a_5 + 2a_6. \qquad (21)$$

To assess the accuracy of these methods, we use the expression for relative error (16).

Table 4 contains the average values for 26 rates from 0.1 c to 0.999 c of the modulus of relative error (16) for the LQZ method and its modifications LQZ$_{m2}$ and LQZ$_{m3}$.

**Table 4**

Average value of the modulus of relative error $\delta X$, %.

| Z | 6 | 12 | 26 | 36 | 52 | 60 |
|---|---|---|---|---|---|---|
| $\langle|\delta X_{LQZ}|\rangle$ | 1.71·10$^{-3}$ | 1.43·10$^{-2}$ | 2.24·10$^{-2}$ | 1.92·10$^{-2}$ | 8.40·10$^{-2}$ | 0.103 |
| $\langle|\delta X_{LQZm2}|\rangle$ | 2.54·10$^{-4}$ | 4.93·10$^{-3}$ | 1.73·10$^{-2}$ | 6.74·10$^{-3}$ | 3.37·10$^{-2}$ | 4.57·10$^{-2}$ |
| $\langle|\delta X_{LQZm3}|\rangle$ | 3.06·10$^{-4}$ | 8.57·10$^{-3}$ | **5.07·10$^{-2}$** | **4.36·10$^{-2}$** | 3.97·10$^{-2}$ | 8.52·10$^{-2}$ |
| Z | 75 | 80 | 92 | 104 | 114 | 118 |
| $\langle|\delta X_{LQZm2}|\rangle$ | 0.105 | 0.102 | 8.15·10$^{-2}$ | 0.145 | 0.267 | 0.332 |
| $\langle|\delta X_{LQZm2}|\rangle$ | 4.91·10$^{-2}$ | 4.49·10$^{-2}$ | 2.86·10$^{-2}$ | 3.13·10$^{-2}$ | 6.23·10$^{-2}$ | 8.55·10$^{-2}$ |
| $\langle|\delta X_{LQZm3}|\rangle$ | **0.141** | **0.146** | **0.128** | 6.83·10$^{-2}$ | 2.48·10$^{-2}$ | 4.58·10$^{-2}$ |

From Table 4, it can be seen that for all the cases considered, the accuracy of the LQZ$_{m2}$-method is higher than the accuracy of the conventional LQZ method. For $Z = 26$, 36, 75, 80, and 92, the accuracy of the LQZ$_{m3}$-method is lower than the accuracy of the usual LQZ-method (highlighted in bold). For $Z = 114$ and $Z = 118$, the accuracy of the LQZ$_{m3}$-method is higher than that of the LQZ$_{m2}$ modified method.

## 3.3 Comparison of accuracy of the LQZ method and its modifications for calculating the ELS of elements from uranium to ohaneson

In [Kats *et al.*, 2022] the LQZ$_{m2}$-method was developed for calculating NMCS, the accuracy of which is generally higher than that of the usual LQZ method.

In Appendix B is an another LQZ$_{m3}$-method for calculating NMCS presented. It is shown that for the copernicium the $\langle ER \rangle$ of LQZ$_{m3}$-method is lower than the $\langle ER \rangle$ of LQZ$_{m2}$-and LQZ-methods, mainly due to the region of low velocities, small and large scattering angles.

In this Section, relative errors $\delta X$ (16) for the LQZ-, LQZ$_{m2}$-, and LQZ$_{m3}$-methods, as well as the arithmetic mean of their modulus depending on $Z$ are calculated in the entire range of $Z$ from 92 to 118 (see Figure 7).

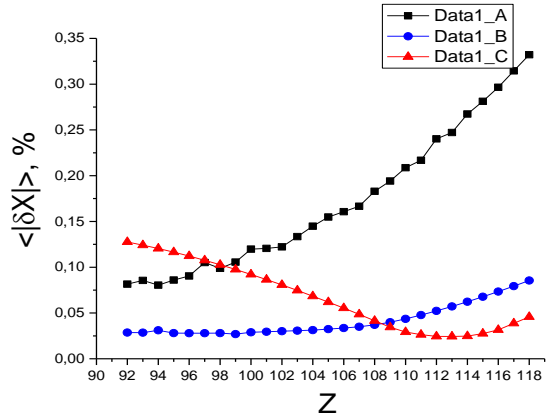

**Fig. 7.** Arithmetic mean values of the modulus of relative error (16) for (A) the LQZ-, (B) LQZ$_{m2}$-, and (C) LQZ$_{m3}$-methods.

From the Figure 7 it can seen that the $\langle |\delta X| \rangle$ error for the LQZ$_{m3}$ method becomes lower than for the LQZ method starting from $Z = 99$, and it is also lower than the error of for LQZ$_{m2}$ - method starting from $Z = 109$.

In Section 3.1, an analytical expression (14) for $X_{LQZ}$ was given, and in Section 3.2, the expressions for $X_{LQZm2}$ (20) and $X_{LQZm3}$ (21) were obtained. In this Section, as before, we will use for comparison the results of [Voskresenskaya, 2017] as reference values, taking into account terms with numbers from 0 to 5000.

Figure 8 shows the $X(\beta)$ dependence for $Z = 114$.

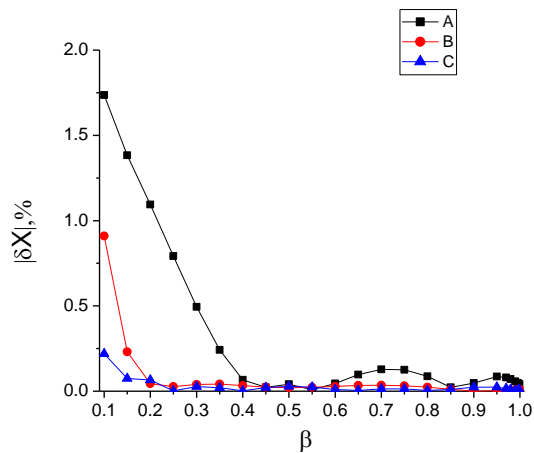

Fig. 8. Modulus of relative error $\delta X$ calculated for (A) the LQZ-, (B) LQZ$_{m2}$-, and (C) LQZ$_{m3}$ - methods at $Z = 114$.

The main contribution to the average error comes from low velocities. Note that the $X$ value is close to 1 for small $\beta$. For example, $X_V = 1.045$ for $Z = 114$ and $\beta = 0.1$, while $X_V = 2.79$ for $\beta = 0.999$.

# 4 Shortened m$a$Ld-LQZS methods for the ELS calculation due to close collisions

### 4.1 Shortened 2$a$3d- and 3$a$3d-LQZS methods for $Z = -118$

As shown in Section 3, the LQZ-method can have significantly higher accuracy for $Z < 0$ in NMCS calculations than for $Z > 0$. This allows us to develop some simplified versions of the LQZ method, such as the so-called shortened LQZ (LQZS) methods for $Z < 0$, which lose only slightly in accuracy compared to the LQZ-method.

In [Kats and Kudravets, 2022] and other works of the authors ([Kudravets, oct 2022, 2023]), were developed a number of the shortened versions of the LQZ method (m$a$Ld-LQZS versions), in which not 30 fitting coefficients are used to approximate the NMCS, but a smaller number of them, from 18 to 6. Reducing the number of coefficients leads, as a rule, to a decrease in the NMCS accuracy, but its accuracy still remains quite high. Previously shortened 3$a$6d-, 3$a$5d-, 3$a$4d-, 2$a$6d-, 2$a$5d-, 2$a$4d-LQZS versions were considered in [Kats and Kudravets, 2022] for $Z = -118$. In this Section, we present the values of the coefficients for the 3$a$3d- and 2$a$3d-LQZS methods and analyze their accuracy on the example of $Z = -118$ (Table 5).

**Table 5**

Coefficients $d_Z$ of shortened 3$a$3d-LQZS (1) and 2$a$3d-LQZS (2) methods for $Z = -118$.

1. 3$a$3d-LQZS method

| j/k | 1 | 2 | 3 |
|---|---|---|---|
| 1 | −0.254322 | −0.783113 | −0.617161 |
| 2 | −6.64629·10⁻³ | −0.053661 | −0.063544 |
| 3 | −9.41278·10⁻³ | −0.065948 | −0.190301 |

2. 2$a$3d-LQZS method

| j/k | 1 | 2 | 3 |
|---|---|---|---|
| 1 | −0.263251 | −0.720550 | −0.436628 |
| 2 | 0.012568 | −0.188279 | −0.451999 |

In the LQZS methods, $a_0$ is replaced by 1. Then, for LQZS method with three coefficients $a_j$, we obtain

$$X_{LQZS_{3aLd}} = 1 + \frac{2\sqrt{2}}{3} a_1 + a_2 + \frac{4\sqrt{2}}{5} a_3.$$

For LQZS method with 2 coefficients $a_j$ $X_{LQZS_{2aLd}}$ reads:

$$X_{LQZS_{2aLd}} = 1 + \frac{2\sqrt{2}}{3}a_1 + a_2.$$

Coefficients $a_j$ are calculated by the formula:

$$a_j(Z,\beta) = \sum_{k=1}^{L} d_Z(j,k)(\beta - \overline{\beta})^{k-1}.$$

$\overline{\beta}$ is somewhat different in the LQZ method and its modified versions (see Section 3).

Table 6 demonstrates the values of the velocity-averaged relative error of the NMCS [Lijian et al., 1995] and the average value of the modulus of the relative error of $X$:

$$\delta X_i = \frac{X(\beta_i) - X_V(\beta_i)}{X_V(\beta_i)}.$$

**Table 6**

Velocity-averaged relative errors of NMCS (1) and the average value of the relative error modulus $X$ (2).

1. Average NMCS values

| maLd-LQZ | LQZ | 3a6d | 3a5d | 2a6d | 2a5d | 3a4d | 2a4d | 3a3d | 2a3d |
|---|---|---|---|---|---|---|---|---|---|
| $\langle ER \rangle$,% | 2.18·10⁻² | 2.22·10⁻² | 2.82·10⁻² | 0.230 | 0.233 | 0.304 | 0.428 | 1.50 | 1.54 |

2. Average $X$ values

| maLd-LQZ | 3a6d | 3a5d | LQZ | 2a5d | 2a6d | 2a4d | 3a4d | 2a3d | 3a3d |
|---|---|---|---|---|---|---|---|---|---|
| $\langle |\delta X| \rangle$,% | 1.03·10⁻² | 1.22·10⁻² | 1.23·10⁻² | 6.00·10⁻² | 6.02·10⁻² | 0.174 | 0.191 | 1.16 | 1.18 |

The error of the LQZS-methods increases sharply when passing from 15 to a smaller number of coefficients $d_z$ that is from 3a5d- to 2a6d-, and 2a5d-LQZS methods. The 2a6d- and 2a5d-LQZS methods give for the considered quantities a smaller average error than the 3a4d-LQZS-method, despite the same or fewer $d_Z$ coefficients. The 2a4d-LQZS method is more accurate than the 3a3d-LQZS method, despite the smaller number of coefficients. It was shown in [Kats and Kudravers, oct 2022] that the 2a4d-LQZS method is more accurate than the 3a3d-LQZS method by the criterion $\langle ER \rangle$ starting from $Z = 59$.

It is interesting that the accuracy of calculating $X$ by the 3a6d- and 3a5d-LQZS methods is higher than the accuracy of the standard LQZ-method. The difference for the 3a5d-LQZS method is not significant and may change when choosing a different set of velocities, but for the 3a6d-LQZS method the difference is noticeable. The accuracy of the 2a6d- and 2a5d-LQZS methods is practically the same.

Surprisingly, the accuracy of the 2*a*4d-LQZS method is higher than the accuracy of the 3*a*4d-LQZS method, and the accuracy of the modified 2*a*3d-LQZS method is higher than the accuracy of the 3*a*3d-LQZS version.

Figure 9 presents the dependence of the $X$ value on $β$.

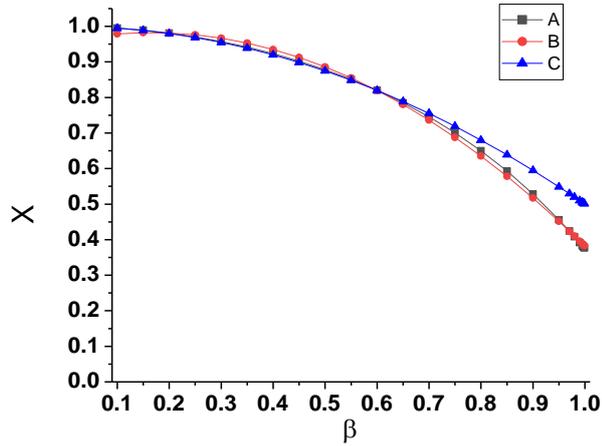

**Fig. 9.** Dependence of $X$ on $β$ for $Z = -118$: numerical calculation [Voskresenskaya, 2017] (A); calculation by the 2*a*3d-LQZS method (B); and the first Born approximation (C).

As was shown in 3 Section, for negative $Z$, the error of the Born approximation for $X$ is small compared to positive $Z$. For $Z = -118$ at $β ≤ 0.6$, the error of the first Born approximation is generally smaller than that in the case of the 3*a*3d- and 2*a*3d-LQZS methods.

### 4.2 Shortened 2aLd- and 3aLd-LQZS methods (L = 3−6) for Z = –60

We will consider in this section ELS (2) due to close collisions of a relativistic bare point nucleus with electrons with $X$ (4). An expression for $X$ in the form of one or another rapidly converging series was obtained in [Voskresenskaya, 2017] and [Lindhard and Sørensen, 1996].

Section 3 shows that the application of the approximate LQZ method [Lijian *et al*, 1995] leads to very high accuracy in calculating the energy straggling for $Z < 0$.

According to the results of the second section, the average $X$ value for the considered set of atomic numbers Z has a maximum at Z =−60. In this section, the $d_Z$ coefficients at $Z = -60$ are calculated for those LQZS-methods for which they were not calculated earlier (Tables 7, 8).

**Table 7**

Coefficients $d_Z$ of shortened $3a$L$d$-LQZS methods (L = 4−6) for $Z = -60$.

3$a$6d-LQZS method

| j/k | 1 | 2 | 3 | 4 | 5 | 6 |
|---|---|---|---|---|---|---|
| 1 | −0.264170 | −0.770877 | −0.409473 | 0.361493 | $-4.23784 \cdot 10^{-3}$ | −0.327724 |
| 2 | $-1.38091 \cdot 10^{-4}$ | −0.046002 | −0.100691 | −0.018124 | −0.023755 | −0.108990 |
| 3 | $-8.20906 \cdot 10^{-4}$ | −0.062825 | −0.267909 | −0.271952 | 0.194871 | 0.313642 |

3$a$5d-LQZS method

| j/k | 1 | 2 | 3 | 4 | 5 |
|---|---|---|---|---|---|
| 1 | −0.263830 | −0.769753 | −0.432726 | 0.320308 | 0.180106 |
| 2 | $-2.50773 \cdot 10^{-5}$ | −0.045628 | −0.108424 | −0.031821 | 0.037551 |
| 3 | $-1.14613 \cdot 10^{-3}$ | −0.063902 | −0.245655 | −0.232536 | 0.018448 |

3$a$4d-LQZS method

| j/k | 1 | 2 | 3 | 4 |
|---|---|---|---|---|
| 1 | −0.264266 | −0.762711 | −0.408683 | 0.243166 |
| 2 | $-1.16014 \cdot 10^{-4}$ | −0.044160 | −0.103411 | −0.047904 |
| 3 | $-1.19080 \cdot 10^{-3}$ | −0.06318 | −0.243192 | −0.240438 |

**Table 8**

Coefficients $d_Z$ of shortened $3a$L$d$-LQZS methods (L = 5−6) for $Z = -60$.

2$a$6d-LQZS method

| j/k | 1 | 2 | 3 | 4 | 5 | 6 |
|---|---|---|---|---|---|---|
| 1 | −0.264170 | −0.711277 | −0.155473 | 0.619486 | −0.189106 | −0.62567 |
| 2 | $-1.38091 \cdot 10^{-3}$ | −0.174246 | −0.647566 | −0.573252 | −0.374029 | −0.513239 |

2$a$5d-LQZS method

| j/k | 1 | 2 | 3 | 4 | 5 |
|---|---|---|---|---|---|
| 1 | −0.262743 | −0.709131 | −0.199681 | 0.540909 | 0.162605 |
| 2 | $-2.36463 \cdot 10^{-3}$ | −0.176068 | −0.609873 | −0.506491 | 0.075209 |

The accuracy of the LQZ-method for calculating the NMCS in the original work [Lijian *et al,* 1995] is characterized by the relative error *ER*. Table 9 show the values of this relative error averaged over the velocities (1) and the average modulus of the relative error δ*X* for a number of LQZS methods and the LQZ method itself (2).

**Table 9**

Velocity-averaged relative error of the NMCS (1) and the average modulus of relative error δ*X* (2).

1. $\langle ER \rangle$,%

| m$a$L$d$/LQZ | LQZ | 3$a$6d | 3$a$5d | 3$a$6d | 2$a$6d | 2$a$5d | 2$a$4d | 3$a$3d | 2$a$3d |
|---|---|---|---|---|---|---|---|---|---|
| $Z = -60$ | $2.33 \cdot 10^{-2}$ | $5.71 \cdot 10^{-2}$ | $6.04 \cdot 10^{-2}$ | 0.128 | 0.317 | 0.319 | 0.366 | 0.398 | 0.591 |
| $Z = -118$ | $2.18 \cdot 10^{-2}$ | $2.22 \cdot 10^{-2}$ | $2.82 \cdot 10^{-2}$ | 0.304 | 0.230 | 0.233 | 0.428 | 1.50 | 1.54 |

2. $\langle |\delta X| \rangle$,%

| m$a$L$d$/LQZ | 3$a$6d | LQZ | 3$a$5d | 2$a$5d | 2$a$6d | 2$a$4d | 3$a$4d | 2$a$3d | 3$a$3d |
|---|---|---|---|---|---|---|---|---|---|
| $Z = -60$ | $1.08 \cdot 10^{-2}$ | $1.39 \cdot 10^{-2}$ | $1.41 \cdot 10^{-2}$ | $7.86 \cdot 10^{-2}$ | $7.93 \cdot 10^{-2}$ | $7.93 \cdot 10^{-2}$ | 0.143 | 0.165 | 0.166 |
| $Z = -118$ | $1.03 \cdot 10^{-2}$ | $1.23 \cdot 10^{-2}$ | $1.22 \cdot 10^{-2}$ | $6.02 \cdot 10^{-2}$ | $6.00 \cdot 10^{-2}$ | 0.191 | 0.174 | 1.18 | 1.16 |

Just as for $Z = -118$, the accuracy of the 3$a$6d-LQZS method for the *X* value calculating is higher than the accuracy of the usual LQZ method. The error of the LQZS methods increases sharply when moving from 15 to a smaller number of the $d_Z$ coefficients. The 2$a$6d- and 2$a$5d-LQZS methods give a smaller average error for the considered quantities than the 3a4d-LQZS method, although the error $\langle ER \rangle$ for the 3a4d-LQZS shortener variant is smaller than for the 2a6d- and 2a5d-LQZS variants. The 2a4d-LQZS method is more accurate than the shortened 3a3d-LQZS method.

The accuracy of the maLd methods increases for a fixed $L$ and an increase in $m$ at $Z = -60$, in contrast to $Z = -118$, although the difference in accuracy between the 3a3d- and 2a3d-LQZS methods is negligible.

For $Z = -60$ and $\beta \leq 0.5$, the error of the first Born approximation for $X$ is, on the whole, less than the error of the 3a3d- and 2a3d-LQZS methods.

The error of the 3a3d- and 2a3d-LQZS methods in calculating $X$ is less than 0.2% at $Z = -60$, while for $Z = -118$ it exceeds 1%. For 3a6d-, 3a5d-, 2a6d-, and 2a5d-LQZS methods, as well as for the usual LQZ method, the accuracy of calculating $X$ for $Z = -118$ is higher than for $Z = -60$.

The average error of the first Born approximation for $X$ is 11.6%, which is higher than the error in its calculation by all shortened LQZS methods at $Z = -118$. Note that according to (3), $X_B$ does not depend on $Z$.

## 5. Summary and conclusions

1.1 In this paper, based on the formalism of the Lindhard–Sørensen, an rigorous expression for the $X$ value, which is reguired to calculate ELS at intermediate energies, was obtained.
1.2 For the first time, the consistency of the results obtained using the formalism of Lindhard–Sørensen and the strict Voskresenskaya method to calculate the ELS was verified for $Z > 0$. With the use a sufficient number of terms in the series of both methods, the results match up to 7 decimal digits and provide the best agreement with the experiment of Scheidenberger et al. [Scheidenberger *et al.,* 1996].
1.3 It is also shown that for negative $Z$, the both ELS formulas lead to the same results for the ELS of heavy relativistic nuclei due to large momentum transfers at moderate relativistic energies.
1.4 Thus, we can conclude that at moderate relativistic energies, when a heavy ion can be considered as a pointlike particles, both methods can be succsessfully used for calculating the energy loss straggling of heavy relativistic ions.
1.5 It was found thah the ELS value taking into account the exact Mott scattering cross-section can be more than 5 times higher than the result obtained in the first Born approximation.

1.6 It was also shown, that the relative correction to the first Born approximation for ELS is positive for $Z > 0$, while it can be positive and negative for $Z < 0$. The relative difference in the ELS taking into account the exact Mott cross section and using the first Born approximation is small for $Z < 0$, for $Z > 0$ it can be an increasing function of $Z$.

2.1 An analytical formula is obtained for calculating the $X$ value in the framework of the approximate LQZ method at moderate relativistic energies.

2.2 It is shown that the velocity-averaged error of ELS calculating by the LQZ method is less than one percent, and for negative $Z$ it does not exceed one hundredth of a percent in order of magnitude.

2.3 This paper proposes a thrice modified LQZ method ($LQZ_{m3}$ method) for calculating NMCS and ELS.

2.4 Analytic expressions are obtained for the ELS calculation of nuclei within the twice and thrice modified ($LQZ_{m2}$ and $LQZ_{m3}$) methods.

2.5 Arithmetic mean values of the modulus of relative error $\delta X$ for methods LQZ, $LQZ_{m2}$, and $LQZ_{m3}$ are estimated in the entire range of $Z$ from 92 to 118.

2.6 It is shown, that for all the cases considered, the accuracy of the $LQZ_{m2}$-method is higher than the accuracy of the usual LQZ method.

3.1 Arithmetic mean of the modulus of relative error $\delta X$ of the $LQZ_{m3}$-method becomes lower than that of the LQZ method starting from $Z = 99$ and lower than that of $LQZ_{m2}$-method starting from $Z = 109$.

3.2 Coefficients for the shortened $3a$3d-, $2a$3d-LQZS methods at $Z = -118$, as well as for the $3a$6d-, $3a$5d-, $3a$4d-, $2a$6d-, and $2a$5d-LQZS methods are estimated at $Z = -60$

3.3 Formulas for calculating the $X$ value by various shortened LQZS methods are obtained.

3.4 Arithmetic mean values of the $\delta X_{LQZS}$ modulus are also calculated for different LQZS-methods.

3.5 It is shown that the accuracy of the m$a$Ld-LQZS methods grows at fixed L and an increase in $m$ for $Z = -60$ (in contrast to $Z = -118$), although the difference in the accuracy the of $2a$3d- and $3a$3d-LQZS methods is insignificant.

3.6 The accuracy of the $3a$6d-LQZS methods at $Z = -60$ is higher than the accuracy of the usual LQZ method, as well as the accuracy of the $3a$5d- and $3a$6d-LQZS methods at $Z = -118$.

**CRediT authorship contribution statement**

**P.B. Kats:** Conceptualization, Methodology, Software, Validation, Writing-Reviewing and Editing; **A.V. Kudravets**: Investigation, Visualization; **A.S. Rymasheuskaya**: Investigation, Visualization; **O.O. Voskresenskaya:** Conceptualization, Methodology, Writing-original draft.

**Declaration of competing interests**

The authors declare that they have no known competing financial interests or personal relationships that could have appeared to influence the work reported in this paper.

## Appendix A: Derivation of the formula for *X* in the framework of Lindhard−Sørensen method

Based on the formula for the transport cross section

$$\sigma_{tr} = 4\pi \lambdabar^2 \sum_{k=-\infty}^{\infty} |k|[(\frac{k-1}{2k-1}\sin^2(\delta_k - \delta_{k-1}) + \frac{1}{2(4k^2-1)}\sin^2(\delta_k - \delta_{-k})], \quad (A.1)$$

where $\lambdabar = \hbar/p$ is the reduced de Broglie wavelength, and on the expression for the fluctuation cross section

$$Q = 2\sigma_{tr} - 4\pi\lambdabar^2 \sum_{k=-\infty}^{\infty} |k|[\frac{(k-1)(k-2)}{(2k-1)(2k-3)}\sin^2(\delta_k - \delta_{k-2})$$
$$+ \frac{k-1}{(2k-3)(4k^2-1)}\sin^2(\delta_k - \delta_{-k+1}) \quad (A.2)$$
$$+ \frac{k+1}{2(2k+1)}\left(\frac{1}{4k^2-1} + \frac{1}{4(k+1)^2-1}\right)\sin^2(\delta_k - \delta_{-k-1})],$$

after substitution (A1) in (A2), we obtain the formula

$$Q = 8\pi\lambdabar^2 \sum_{k=-\infty}^{\infty} |k|[\frac{k-1}{2k-1}\sin^2(\delta_k - \delta_{k-1}) + \frac{1}{2(4k^2-1)}\sin^2(\delta_k - \delta_{-k})$$
$$- \frac{(k-1)(k-2)}{2(2k-1)(2k-3)}\sin^2(\delta_k - \delta_{k-2}) - \frac{k-1}{2(2k-3)(4k^2-1)}\sin^2(\delta_k - \delta_{-k+1}) - \quad (A.3)$$
$$- \frac{k+1}{4(2k+1)}\left(\frac{1}{4k^2-1} + \frac{1}{4(k+1)^2-1}\right)\sin^2(\delta_k - \delta_{-k-1})],$$

which is connected with the relation for increase in the energy loss:

$$\frac{dW}{dx} = nZ_2 \frac{E_0^2}{4} Q \quad (A.4)$$

[Lindhard and Sørensen, 1996].

From the formula (2) of the second Section, we can represent the *X* value as

$$X = \frac{E_0^2 Q(1-\beta^2)}{16\pi Z^2 e^4} \tag{A.5}$$

and, taking into account the formula (11), obtain it in the form:

$$X = \frac{m^2 c^4 \beta^4}{4\pi Z^2 e^4 (1-\beta^2)} Q. \tag{A.6}$$

Now we find the product of the coefficients from (A2) and (A6):

$$8\pi \lambdabar^2 \frac{m^2 c^4 \beta^4}{4\pi Z^2 e^4 (1-\beta^2)} = \frac{2 m^2 c^4 \beta^4 \lambdabar^2}{Z^2 e^4 (1-\beta^2)} = 2\frac{\hbar^2 c^2 \beta^2}{Z^2 e^4} = \frac{2}{\eta^2}, \tag{A.7}$$

where $\eta = Z\alpha/\beta$.

Thus, we finally obtain a rigorous formula for calculating the $X$ value (4) in the form of convergent series:

$$X = \frac{2}{\eta^2} \sum_{k=-\infty}^{\infty} |k| \Big[ \frac{k-1}{2k-1} \sin^2(\delta_k - \delta_{k-1}) + \frac{1}{2(4k^2-1)} \sin^2(\delta_k - \delta_{-k}) -$$

$$- \frac{(k-1)(k-2)}{2(2k-1)(2k-3)} \sin^2(\delta_k - \delta_{k-2}) - \frac{k-1}{2(2k-3)(4k^2-1)} \sin^2(\delta_k - \delta_{-k+1}) -$$

$$- \frac{k+1}{4(2k+1)} \Big( \frac{1}{4k^2-1} + \frac{1}{4(k+1)^2-1} \Big) \sin^2(\delta_k - \delta_{-k-1}) \Big].$$

## Appendix B: Thrice modified Lijian−Qing−Zhengming Method

Consider a modification of the LQZ method, which can be called a thrice modified LQZ method (LQZ$_{m3}$). The LQZ$_{m3}$-method can be represented by the following expressions:

$$R_{LQZm3}(\theta, Z, E) = 1 + \sum_{j=1}^{6} a_j(Z, E)(1-\cos\theta)^{j/2},$$

$$a_j(Z, E) = \sum_{k=1}^{5} d_Z(j,k)(\beta - \bar{\beta})^{k-1}, \bar{\beta} = 0.668269, \tag{B.1}$$

where $\bar{\beta}$ is the arithmetic mean for 26 relative speeds from 0.1 to 0.999. The number of coefficients for each element remains the same as in the LQZ- and LQZ$_{m2}$-methods. The hypothesis is that due to the larger number of coefficients, the accuracy of predicting the angular dependence of the scattering cross section will increase.

Let us consider the method of calculating the coefficients $d_Z(j,k)$. At the first stage, based on the least squares method, 26 systems of equations are compiled to find the coefficients $a_j(Z, \beta_i), j = 1-6, i = 1, ..., 26$ (see below).

$$\sum_{l=1}^{36}(1-\cos\theta_l)^{1/2} + \sum_{l=1}^{36}a_1(Z,\beta_i)(1-\cos\theta_l)^{2/2} + \sum_{l=1}^{36}a_2(Z,\beta_i)(1-\cos\theta_l)^{3/2} + \sum_{l=1}^{36}a_3(Z,\beta_i)(1-\cos\theta_l)^{4/2} +$$
$$\sum_{l=1}^{36}a_4(Z,\beta_i)(1-\cos\theta_l)^{5/2} + \sum_{l=1}^{36}a_5(Z,\beta_i)(1-\cos\theta_l)^{6/2} + \sum_{l=1}^{36}a_6(Z,\beta_i)(1-\cos\theta_l)^{7/2} =$$
$$= \sum_{l=1}^{36}R_{KHV}(\theta_l;Z,\beta_i)(1-\cos\theta_l)^{1/2};$$

$$\sum_{l=1}^{36}(1-\cos\theta_l)^{2/2} + \sum_{l=1}^{36}a_1(Z,\beta_i)(1-\cos\theta_l)^{3/2} + \sum_{l=1}^{36}a_2(Z,\beta_i)(1-\cos\theta_l)^{4/2} + \sum_{l=1}^{36}a_3(Z,\beta_i)(1-\cos\theta_l)^{5/2} +$$
$$\sum_{l=1}^{36}a_4(Z,\beta_i)(1-\cos\theta_l)^{6/2} + \sum_{l=1}^{36}a_5(Z,\beta_i)(1-\cos\theta_l)^{7/2} + \sum_{l=1}^{36}a_6(Z,\beta_i)(1-\cos\theta_l)^{8/2} =$$
$$\sum_{l=1}^{36}R_{KHV}(\theta_l;Z,\beta_i)(1-\cos\theta_l)^{2/2};$$

$$\sum_{l=1}^{36}(1-\cos\theta_l)^{3/2} + \sum_{l=1}^{36}a_1(Z,\beta_i)(1-\cos\theta_l)^{4/2} + \sum_{l=1}^{36}a_2(Z,\beta_i)(1-\cos\theta_l)^{5/2} + \sum_{l=1}^{36}a_3(Z,\beta_i)(1-\cos\theta_l)^{6/2} +$$
$$\sum_{l=1}^{36}a_4(Z,\beta_i)(1-\cos\theta_l)^{7/2} + \sum_{l=1}^{36}a_5(Z,\beta_i)(1-\cos\theta_l)^{8/2} + \sum_{l=1}^{36}a_6(Z,\beta_i)(1-\cos\theta_l)^{9/2} =$$
$$= \sum_{l=1}^{36}R_{KHV}(\theta_l;Z,\beta_i)(1-\cos\theta_l)^{3/2};$$

$$\sum_{l=1}^{36}(1-\cos\theta_l)^{4/2} + \sum_{l=1}^{36}a_1(Z,\beta_i)(1-\cos\theta_l)^{5/2} + \sum_{l=1}^{36}a_2(Z,\beta_i)(1-\cos\theta_l)^{6/2} + \sum_{l=1}^{36}a_3(Z,\beta_i)(1-\cos\theta_l)^{7/2} +$$
$$\sum_{l=1}^{36}a_4(Z,\beta_i)(1-\cos\theta_l)^{8/2} + \sum_{l=1}^{36}a_5(Z,\beta_i)(1-\cos\theta_l)^{9/2} + \sum_{l=1}^{36}a_6(Z,\beta_i)(1-\cos\theta_l)^{10/2} =$$
$$= \sum_{l=1}^{36}R_{KHV}(\theta_l;Z,\beta_i)(1-\cos\theta_l)^{4/2};$$

$$\sum_{l=1}^{36}(1-\cos\theta_l)^{5/2} + \sum_{l=1}^{36}a_1(Z,\beta_i)(1-\cos\theta_l)^{6/2} + \sum_{l=1}^{36}a_2(Z,\beta_i)(1-\cos\theta_l)^{7/2} + \sum_{l=1}^{36}a_3(Z,\beta_i)(1-\cos\theta_l)^{8/2} +$$
$$\sum_{l=1}^{36}a_4(Z,\beta_i)(1-\cos\theta_l)^{9/2} + \sum_{l=1}^{36}a_5(Z,\beta_i)(1-\cos\theta_l)^{10/2} + \sum_{l=1}^{36}a_6(Z,\beta_i)(1-\cos\theta_l)^{11/2} =$$
$$= \sum_{l=1}^{36}R_{KHV}(\theta_l;Z,\beta_i)(1-\cos\theta_l)^{5/2};$$

$$\sum_{l=1}^{36}(1-\cos\theta_l)^{6/2}+\sum_{l=1}^{36}a_1(Z,\beta_i)(1-\cos\theta_l)^{7/2}+\sum_{l=1}^{36}a_2(Z,\beta_i)(1-\cos\theta_l)^{8/2}+\sum_{l=1}^{36}a_3(Z,\beta_i)(1-\cos\theta_l)^{9/2}+$$

$$\sum_{l=1}^{36}a_4(Z,\beta_i)(1-\cos\theta_l)^{10/2}+\sum_{l=1}^{36}a_5(Z,\beta_i)(1-\cos\theta_l)^{11/2}+\sum_{l=1}^{36}a_6(Z,\beta_i)(1-\cos\theta_l)^{12/2}=$$

$$=\sum_{l=1}^{36}R_{KHV}(\theta_l;Z,\beta_i)(1-\cos\theta_l)^{6/2},\,\theta_l=5l°.$$

(B2)

Here, $R_{KHV}(\theta;Z,\beta)$ is the normalized Mott scattering cross section calculated by the method considered in [Kats et al., 2022].

At the second stage, based on the least squares, 5 systems ($j = 1–5$) of equations are compiled to find the coefficients $d_Z(j,k)$:

$$\sum_{i=1}^{26}[d_Z(j,1)+d_Z(j,2)(\beta_i-\overline{\beta})+d_Z(j,3)(\beta_i-\overline{\beta})^2+d_Z(j,4)(\beta_i-\overline{\beta})^3+d_Z(j,5)(\beta_i-\overline{\beta})^4]=$$

$$=\sum_{i=1}^{26}a_j(Z,\beta_i);$$

$$\sum_{i=1}^{26}[d_Z(j,1)(\beta_i-\overline{\beta})+d_Z(j,2)(\beta_i-\overline{\beta})^2+d_Z(j,3)(\beta_i-\overline{\beta})^3+d_Z(j,4)(\beta_i-\overline{\beta})^4+$$

$$+d_Z(j,5)(\beta_i-\overline{\beta})^5]=\sum_{i=1}^{26}a_j(Z,\beta_i)(\beta_i-\overline{\beta});$$

$$\sum_{i=1}^{26}[d_Z(j,1)(\beta_i-\overline{\beta})^2+d_Z(j,2)(\beta_i-\overline{\beta})^3+d_Z(j,3)(\beta_i-\overline{\beta})^4+d_Z(j,4)(\beta_i-\overline{\beta})^5+$$

$$+d_Z(j,5)(\beta_i-\overline{\beta})^6]=\sum_{i=1}^{26}a_j(Z,\beta_i)(\beta_i-\overline{\beta})^2;$$

$$\sum_{i=1}^{26}[d_Z(j,1)(\beta_i-\overline{\beta})^3+d_Z(j,2)(\beta_i-\overline{\beta})^4+d_Z(j,3)(\beta_i-\overline{\beta})^5+d_Z(j,4)(\beta_i-\overline{\beta})^6+$$

$$+d_Z(j,5)(\beta_i-\overline{\beta})^7]=\sum_{i=1}^{26}a_j(Z,\beta_i)(\beta_i-\overline{\beta})^3;$$

$$\sum_{i=1}^{26}[d_Z(j,1)(\beta_i-\overline{\beta})^4+d_Z(j,2)(\beta_i-\overline{\beta})^5+d_Z(j,3)(\beta_i-\overline{\beta})^6+d_Z(j,4)(\beta_i-\overline{\beta})^7+$$

$$d_Z(j,5)(\beta_i-\overline{\beta})^8]=\sum_{i=1}^{26}a_j(Z,\beta_i)(\beta_i-\overline{\beta})^4.$$

(B.3)

As a result, 30 coefficients are determined.

Tables B1 and B2 list the calculated coefficients for copernicium for the LQZ$_{m2}$- and LQZ$_{m3}$-methods, respectively.

**Table B1**

Coefficients $d_Z(j,k)$ of the modified $LQZ_{m2}$-method

| j/k | 1 | 2 | 3 | 4 | 5 | 6 |
|---|---|---|---|---|---|---|
| 1 | 1.794354 | 4.835146 | −28.78201 | −45.41782 | 189.0485 | 209.4923 |
| 2 | −10.21441 | −43.45847 | 187.7235 | 459.5930 | −1373.950 | −1913.595 |
| 3 | 16.29755 | 102.3528 | −341.6067 | −1193.269 | 2966.108 | 4829.543 |
| 4 | −7.197864 | −70.89085 | 249.4377 | 1116.197 | −2475.503 | −4515.852 |
| 5 | 0.543736 | 12.05521 | −67.20368 | −344.3931 | 706.8487 | 1405.938 |

**Table B2**

Coefficients $d_Z(j,k)$ of the modified $LQZ_{m3}$-method

| j/k | 1 | 2 | 3 | 4 | 5 |
|---|---|---|---|---|---|
| 1 | −0.046366 | 7.214773 | 27.33282 | −55.49846 | −185.6567 |
| 2 | 6.346974 | −64.71300 | −318.4370 | 546.0103 | 2008.813 |
| 3 | −33.01762 | 170.3431 | 1126.851 | −1579.952 | −6760.729 |
| 4 | 57.51741 | −169.9717 | −1596.318 | 1894.950 | 9562.924 |
| 5 | −38.29772 | 78.58116 | 982.4118 | −1006.053 | −5998.759 |
| 6 | 8.728996 | 16.66040 | −221.7897 | 195.7517 | 1381.087 |

Accuracy of $R_{LQZ}(\theta; Z, \beta)$ is characterized by the average relative error *ER*:

$$ER(Z,\beta) = \sqrt{\frac{\sum_{i=0}^{36}[R_{LQZ}(\theta_i;Z,\beta) - R_M(\theta_i;Z,\beta)]^2}{\sum_{i=0}^{36} R_M(\theta_i;Z,\beta)^2}}.$$

Figure B1 demonstrates the *ER*($\beta$) dependence for the LQZ-, $LQZ_{m2}$- and $LQZ_{m3}$-methods.

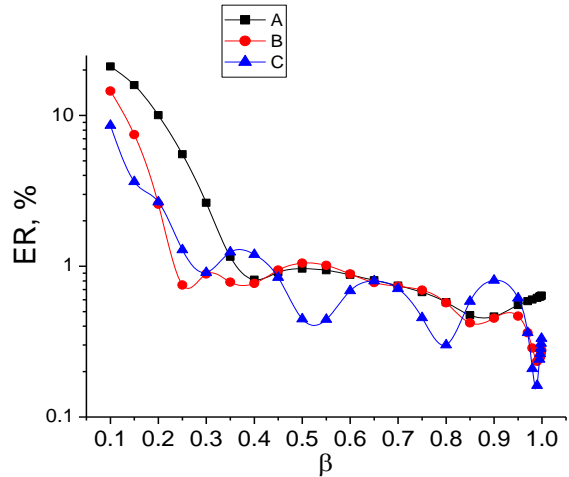

Fig. B1. Relative error *ER* as a function of relative velocity $\beta$ for (A) LQZ-, (B) $LQZ_{m2}$-, and (C) $LQZ_{m3}$-methods.

The accuracy of the $LQZ_{m2}$-method and especially the $LQZ_{m3}$-method is significantly higher than the accuracy of the LQZ-method for $v < 0.3$ c and $v > 0.96$ c. At $\beta = 0.15$, $ER_{LQZ} =$ 15.9%, $ER_{LQZm2} = 7.5\%$, and $ER_{LQZm3} = 2.7\%$. In the range of 0.8–0.9 $c$, the accuracy of the $LQZ_{m3}$-method is lower than that of other methods, but the error for copernicium remains less than 1%.

Figure B2 shows the NMCS values obtained by numerical summation of the Mott series, as well as by the LQZ-, $LQZ_{m2}$- and $LQZ_{m3}$-methods for copernicium.

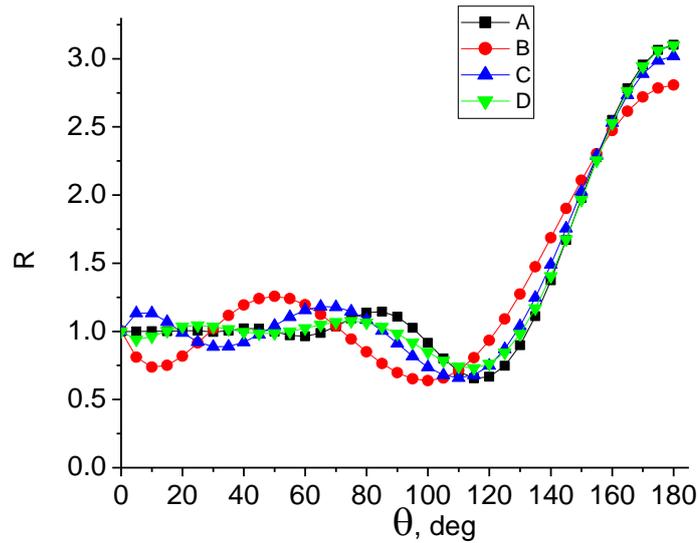

Fig. B2. Dependence of the NMCS on the angle $\theta$ (deg) for $Z = 112$ and $\beta = 0.15$ (A) in the numerical calculation, as well as in the calculation by (B) the LQZ-, (C) $LQZ_{m2}$,- and (D) $LQZ_{m3}$ -methods.

The relative error averaged over the velocities is given in Table B3.

**Table B3**

Velocity-averaged relative error for copernicium

| Method | LQZ | $LQZ_{m2}$ | $LQZ_{m3}$ |
|---|---|---|---|
| $\langle ER \rangle$,% | 2.70 | 1.46 | 1.09 |

For the copernicium, the error $\langle ER \rangle$ for the $LQZ_{m3}$-method is lower than for the $LQZ_{m2}$- and LQZ-methods, mainly due to the region of low velocities, small and large scattering angles.

From Figure B2 it can be seen that the $LQZ_{m3}$-method is in good agreement with the numerical calculation for small and large $\theta$.

In the range of angles where the NMCS, according to numerical calculations, differs little from 1, the LQZ- and $LQZ_{m2}$-methods give noticeable deviations from 1. For backscattering, where the NMCS exceeds 3, the error of the LQZ-, LQZ$_{m2}$,-and LQZ$_{m3}$-methods is about 10%, 3%, and 0.14 %, respectively.